\documentstyle[aps,preprint,epsfig,floats]{revtex}

\begin{document}

\def\d{{\rm d}}
\def\e{{\rm e}}
\def\O{{\rm O}}
\def\half{\mbox{$\frac12$}}
\def\eref#1{(\protect\ref{#1})}
\def\etal{{\it{}et~al.}}
\def\av#1{\left\langle#1\right\rangle}
\def\set#1{\left\lbrace#1\right\rbrace}
\def\from{\leftarrow}

\newdimen\captwidth
\captwidth=5.5in
\def\capt#1{\refstepcounter{figure}\bigskip\hbox to \textwidth{%
       \hfil\vbox{\hsize=\captwidth\renewcommand{\baselinestretch}{1}\small
       {\sc Figure \thefigure}\quad#1}\hfil}\bigskip}

\renewcommand{\topfraction}{0.9}
\renewcommand{\textfraction}{0.1}
\renewcommand{\floatpagefraction}{0.9}
\setlength{\tabcolsep}{4pt}
\setcounter{topnumber}{1}

\title{Clustering and preferential attachment in growing networks}
\author{M. E. J. Newman}
\address{Santa Fe Institute, 1399 Hyde Park Road, Santa Fe, NM 87501}
\maketitle

\vspace{0.8in}
\begin{abstract}
  We study empirically the time evolution of scientific collaboration
  networks in physics and biology.  In these networks, two scientists are
  considered connected if they have coauthored one or more papers together.
  We show that the probability of scientists collaborating increases with
  the number of other collaborators they have in common, and that the
  probability of a particular scientist acquiring new collaborators
  increases with the number of his or her past collaborators.  These
  results provide experimental evidence in favor of previously conjectured
  mechanisms for clustering and power-law degree distributions in networks.
\end{abstract}

\newpage

\section{Introduction}
Many systems take the form of networks---sets of nodes, or vertices, joined
together by links, or edges.  The Internet, the power grid, social
networks, food webs, distribution networks, and metabolic networks are
commonly cited examples.  Investigations of networks within the physics
community fall loosely into two categories: (1)~studies of static network
structure~\cite{WS98,Watts99,AJB99,Broder00,ASBS00,NWS01} and dynamical
processes taking place on fixed networks~\cite{Strogatz01,Monasson99,MN00};
(2)~studies of the dynamics of networks themselves---how and why their
topology changes over time~\cite{WS98,Watts99,BA99,KRL00,DMS00}.  It is
this second category that we address here, focusing on two properties which
have received a large amount of attention in the literature---clustering
and preferential attachment.

Sociologists have long known that social networks---networks of personal
acquaintances, for example---display a high degree of transitivity, meaning
that there is a heightened probability of two people being acquainted if
they have one or more other acquaintances in common.  In the physics
literature this phenomenon is called ``clustering.''  Watts and
Strogatz~\cite{WS98} measured clustering in a number of real-world
networks, including both social and physical networks, by calculating a
clustering coefficient, equal to the probability that two vertices that are
both neighbors of the same third vertex will be neighbors of one another.
They found that in many networks the clustering coefficient is much higher
than its expected baseline value, which is set by comparison with a random
graph.

It has also been pointed out by a number of
authors~\cite{AJB99,Broder00,ASBS00,FFF99}, particularly in studies of the
Internet and the World-Wide Web, that real-world networks have highly
skewed distributions of vertex degree.  (The degree of a vertex is the
number of other vertices to which it is connected.)
In many cases, the degree distribution is found to follow a power law, a
particularly telling functional form which often signifies an underlying
process worthy of study.

Explanations have been put forward for both of these observations.  In the
case of clustering, it is conjectured that pairs of individuals with a
common acquaintance (or several) are likely to become acquainted themselves
through introduction by their mutual friend(s)~\cite{Watts99}.  In the case
of degree distributions, it is conjectured that, for a variety of reasons,
vertices accumulate new edges in proportion to the number they have
already, leading to a multiplicative process which is known to give
power-law distributions~\cite{BA99,KRL00,DMS00}.  This process is often
called ``preferential attachment.''  While both of these explanations are,
in some contexts at least, perfectly plausible, there has been little if
any empirical evidence in their favor---a glaring problem for two
conjectures which have formed the foundation of a substantial body of
research.  The principal reason for this has been the lack of good
time-resolved data on how networks grow.

In order to test a conjecture such as ``people with many common friends are
more likely to become acquainted than those with few or none,'' one needs
to watch a network grow and see if the process described by the conjecture
does indeed happen with significantly heightened frequency.  Although data
on the structure of networks are quite plentiful, data on how they grow
have proved harder to come by.  Recently, however, the author conducted
some empirical studies of collaboration networks of scientists: networks in
which pairs of scientists are linked together if they have coauthored one
or more papers~\cite{Newman01a,Newman01b}.  These collaboration
networks are true social networks, since two scientists who have coauthored
a paper will normally be acquainted with one another.  (There are
occasional exceptions---see Ref.~\onlinecite{Newman01b}.)  They are also
well documented, since there exist extensive machine-readable
bibliographies of the scientific literature.  What's more, as Barab\'asi
and co-workers have recently pointed out~\cite{BJNRSV01}, these networks
have excellent time resolution as well, because each paper comes with a
publication or receipt date.  As we now show, this allows us to test
directly the clustering and preferential attachment conjectures.

In this study we look at collaboration networks derived from two
bibliographic sources:
\begin{enumerate}
\item The Los Alamos E-print Archive, a database of preprints in physics,
  self-submitted by their authors;
\item Medline, a database of published papers in biology and medicine,
  whose entries are professionally maintained by the National Institutes of
  Health.
\end{enumerate}
While neither of these databases records the exact publication date of the
papers they contain, both include a record of the sequence in which papers
were added to the database.  This is enough for our purposes: all that we
need for our calculations is the order of the collaborations undertaken by
each author in the database, and the order of the papers is a reasonable
proxy for this---probably not correct in every case, but assumed to be
correct in most.  Two other databases that we studied
previously~\cite{Newman01a} do not contain enough information to establish
order of collaborations, recording publication or database entry of papers
to the nearest year only.  This creates ambiguity since many authors
produce more than one paper a year, and so we did not use these databases
for the current study.

Authors are identified by their full surname and all initials.  As
discussed previously~\cite{Newman01a,Newman01b}, an author who gives their
name differently on different papers may be confused for two people by this
measure, while two people with identical surnames and initials may be
confused for one.  The error in the number of vertices in the network as a
result of these problems was found to be on the order of~5\%.

We study a six-year interval of time for both databases.  (For the Los
Alamos Archive we use 1995 to 2000 inclusive, for Medline 1994 to 1999.)
Over this period the Los Alamos Archive records $58\,342$ distinct names,
and Medline $1\,648\,660$.  In each of the calculations presented here, we
use the first five of the six years to construct a collaboration network,
and then examine how that network further changes in the remaining one
year.  Our assumption is that any scientist who is currently active will
produce at least one paper during the initial five year period, as will any
currently active collaboration between a pair of scientists, so that the
network we have at the end of that period will be essentially complete.
New vertices added in the sixth year represent, it is assumed, new
individuals entering the field, and new edges represent genuine new
collaborations.  Of course there are some exceptions, such as established
scientists who for one reason or another fail to publish anything for five
years and then produce a paper in the sixth, and these will be
misrepresented in our calculations.  We assume these are a small fraction
of the total.  There will also be some scientists who leave the field
during the six years, to go into different fields or professions, or
because they retire.  We make no attempt to guess which individuals leave
in this way: everyone whose name appears even once is considered a member
of the network for the entire period of study thereafter.  This will
introduce some error into our calculations.  However, it is straightforward
to convince oneself that the correlations we are looking for in the present
study will only be weakened by this error, not strengthened, so there is no
danger of false positive results.

\section{Clustering}
Let us consider first the question of clustering in the network.  We
already know that the clustering coefficient is high in our collaboration
networks---$0.45$ for the Los Alamos Archive and $0.088$ for Medline over a
five-year period~\cite{Newman01a}.  The calculation presented here improves
on these results in two ways.  First, the simple clustering coefficient
includes contributions from collaborations between authors which {\em
  preceded\/} their collaborations with any mutual acquaintances.  By using
time-resolved data we can exclude these collaborations from our measure of
clustering.  Second, we can determine whether the probability of two
individuals collaborating increases as the number $m$ of their previous
mutual acquaintances goes up.  If this is the case, then it suggests that
the standard explanation of clustering---introduction of future
collaborators to one another by common previous acquaintances---is correct,
the probability of such an introduction presumably increasing with~$m$.
Other explanations, such as the institutional explanation proposed in
Ref.~\cite{Newman01b}, would be harder to justify.

Measuring the probability of collaboration between authors as a function of
their number of mutual acquaintances is complicated by the fact that both
the size of the graph and the numbers of mutual acquaintances themselves
are changing over time.  We consider the probability $P_m(t)$ that the two
scientists connected by a link added at time $t$ have $m$ mutual
acquaintances.  (Time is somewhat arbitrary here.  It can be real time, but
it can also be any other function which increases monotonically as papers
are added to the database---only the order of the papers matters, not their
precise timing.  The links created by a paper with three or more authors
are all considered to be added at the same instant.)  We have
\begin{equation}
P_m(t) = {n_m(t)\over \frac12 N(t)[N(t)-1]} R_m,
\label{defspi}
\end{equation}
where $n_m(t)$ is the number of pairs with $m$ mutual acquaintances
immediately {\em before\/} the addition of the paper at time~$t$, $N(t)$ is
the current number of authors in the network, and $R_m$ is the relative
probability of collaboration between the two scientists connected by this
link, i.e.,~the ratio between the actual probability of their collaborating
and the probability of their collaborating in a network in which presence
of mutual acquaintances makes no difference.  We assume that the
probability that two scientists with a given value of $m$ collaborate at a
particular time does not depend on the number of other scientists with that
value of $m$, or on the size of the database in which the paper they write
is archived, and hence that $R_m$ is independent of~$t$~\cite{note1}.  This
makes it a suitable quantity to measure to test our clustering hypothesis.
In a world with no clustering, we would have $R_m=1$ for all~$m$; in a
world in which clustering arises through introductions, as above, it should
increase with increasing~$m$.

\begin{figure}
\begin{center}
\psfig{figure=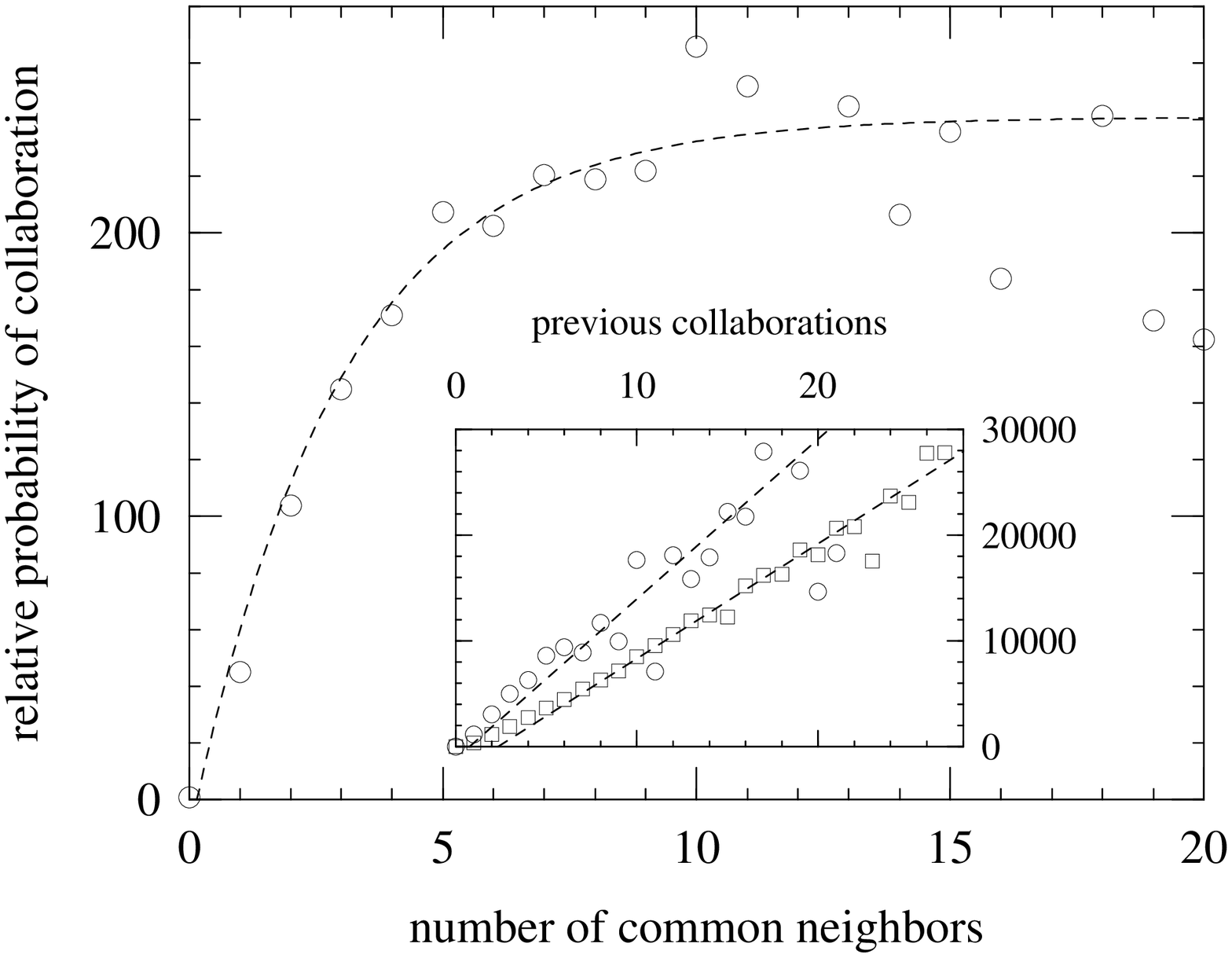,width=4.5in}
\end{center}
\capt{Probability of collaboration between scientists in the Los
  Alamos Archive as a function of their number of mutual previous
  collaborators.  The dotted line is the best fit of the
  form~\eref{expfit}.  Inset: the relative probability of collaboration as
  a function of number of previous collaborations of the same scientists,
  for the Los Alamos Archive (circles) and Medline (squares).  The dotted
  lines are the best straight-line fits to the data.  The data for Medline
  have been divided by a factor of 50 vertically to improve the clarity of
  the figure.}
\label{clustering}
\end{figure}

To measure $R_m$, one simply constructs a histogram of the value of $m$ for
each link added to the graph in which each sample is weighted by a factor
of $\frac12 N(t)[N(t)-1]/n_m(t)$.  In Fig.~\ref{clustering} we do this for
the network of the Los Alamos Archive.  As discussed above, we evaluate
$R_m$ for the last of our six years only, the previous five being used to
establish the initial network for the calculation.  As the figure shows,
$R_m$ does indeed increase with~$m$, and is much greater than~1 for all
$m>0$.  A pair of scientists who have five mutual previous collaborators,
for instance, are about twice as likely to collaborate as a pair with only
two, and about 200 times as likely as a pair with none.  $R_m$~increases
roughly linearly for small $m$, perhaps indicating that each common
collaborator of a pair of scientists is equally likely to introduce them.
The curve appears to flatten off for higher~$m$, although the data become
poor for $m\gtrsim8$, since the number of pairs of authors with this many
common collaborators who have not already collaborated themselves is very
small.

As well as supporting the standard explanation of clustering in social
networks, our data for $R_m$ might prove useful for modeling purposes.  For
example, in some models of the growth of social
networks~\cite{Watts99,JN01}, a particular form is assumed for the
probability of individuals becoming acquainted, as a function of their
number of mutual friends.  Fig.~\ref{clustering} provides a rough empirical
guide for what that functional form should be.  In the figure we give a fit
to the data of the form
\begin{equation}
R_m = A - B \e^{-m/m_0},
\label{expfit}
\end{equation}
where $A$, $B$, and $m_0$ are constants.  This form appears to fit
reasonably well and might be suitable for use in the models.

\section{Repeat collaborations}
In the calculation described above, we included only newly appearing edges
in the network.  Repeat collaborations between authors who had collaborated
before were excluded; we assume that such collaborations are more likely to
be a result of previous acquaintance than the result of network structure.
This however raises another interesting question: does probability of
collaboration also increase with the number of times one has collaborated
before?  The answer is yes, as shown in the inset of Fig.~\ref{clustering},
which measures the relative probability $R_n$ (defined similarly to $R_m$
above) of two coauthors collaborating if they have collaborated $n$ times
previously within the period covered by our study.  If collaboration
probability were independent of previous collaboration, we would have
$R_n=1$ for all $n$, but as the figure shows, $R_n$ increases roughly
linearly with~$n$, indicating that number of past collaborations is a good
indicator of the probability of future collaboration.  However, one must
bear in mind that this calculation may be influenced by varying frequencies
of collaboration: regular collaborators who publish often will have more
publications in the database as well as greater likelihood of publishing
again in the last of our six years, producing a correlation just as seen in
the figure.  To eliminate this effect one would have to look at data for a
longer period of time and compare collaborators with similar numbers of
publications but different publication rates.  Unfortunately, this is not
practical with the data available to us at present.

\section{Preferential attachment}
We can also use our data to test for preferential attachment in the
collaboration network.  Barab\'asi~\etal~\cite{BJNRSV01} have previously
looked for preferential attachment in two collaboration networks derived
from data for publications in mathematics and neuroscience.  Papers in
their databases were dated only to the nearest year, making the order in
which collaborations occur uncertain, as discussed above.  To get around
this, they restricted themselves to measuring the number of new papers each
author in the network published in a single year, as a function of number
of previous papers.  This should be an increasing function if there is
preferential attachment, or constant otherwise.  Their results show a clear
increase and hence favor preferential attachment.

Using our data we can measure preferential attachment in our networks
directly by a method similar to the one we used to measure clustering
above.  We define a relative probability $R_k$ that a link added at time
$t$ connects to a vertex representing a scientist who has collaborated
previously with $k$ others.  By analogy with Eq.~\eref{defspi}, the
corresponding absolute probability $P_k(t)$ that this link connects to a
vertex with degree $k$ is $P_k(t) = R_k n_k(t)/N(t)$, where $n_k(t)$ is the
number of vertices with degree $k$ immediately before addition of this
link.  Then $R_k$ can be estimated by making a histogram of the degrees $k$
of the vertices to which each link is added in which each sample is
weighted by a factor of $N(t)/n_k(t)$.  If there is no preferential
attachment, $R_k$~should equal~1 for all~$k$.  If there is preferential
attachment, it should be an increasing function of~$k$, and the widely held
belief is that it should in fact increase linearly with~$k$.  If it
increases linearly, then the resulting degree distribution of the network
will be a power law~\cite{BA99,KRL00,DMS00}.

\begin{figure}
\begin{center}
\psfig{figure=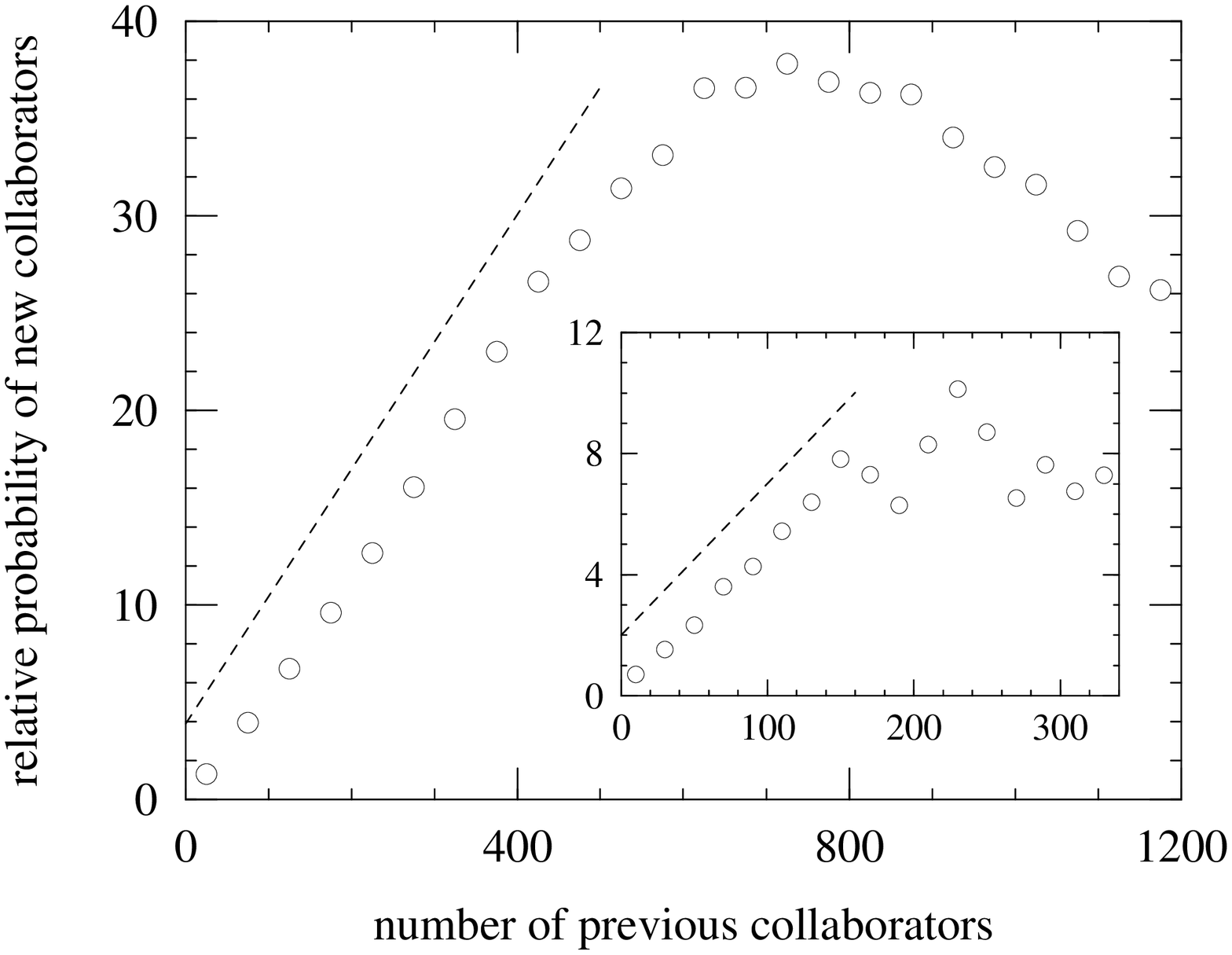,width=4.5in}
\end{center}
\capt{The relative probability that a new edge in the collaboration
  network will connect to a vertex of given degree.  The main figure shows
  data from the Medline database, the inset data from the Los Alamos
  E-print Archive.}
\label{prefer}
\end{figure}

In Fig.~\ref{prefer} and its inset we show empirical results for $R_k$ for
the databases studied here.  As the figure shows, the relative probability
is in both cases close to linear in the initial part of the curve, but
falls off once $k$ becomes large.  This is understandable: no one can
collaborate with an infinite number of people in a finite period of time,
so at some point $R(k)$ must start to decrease.  This point appears to be
around 150 collaborators in physics and 600 in biomedicine.  Interestingly,
these figures coincide roughly with the points at which the observed degree
distribution in these networks starts to deviate from the power-law
form~\cite{Newman01b}, lending support to the theory that preferential
attachment is the origin of the power law.

Our results differ somewhat from those of Barab\'asi~\etal~\cite{BJNRSV01},
who found preferential attachment for their networks, but did not find
linear behavior.  In the language used here, their finding was that
$R(k)\sim k^\nu$, with $\nu\simeq0.8$.  This form does not fit our data
very well.  A power-law fit to the increasing part of $R(k)$ for our data
gives $\nu=1.04\pm0.04$ for Medline and $\nu=0.89\pm0.98$ for the Los
Alamos Archive, both of which are compatible with the conjecture of linear
preferential attachment, while only the latter is compatible with
$\nu=0.8$.  In practice however, this difference may have little effect.
As Krapivsky~\etal~\cite{KRL00} have shown, sub-linear preferential
attachment gives rise to a stretched exponential cutoff in the resulting
degree distribution, but we already have a similar cutoff in our
distribution as a result of the deviation of $R(k)$ from linear behavior
for large enough~$k$.

\section{Conclusions}
To conclude, we have measured the probability of collaboration between
scientists in two collaboration networks as a function of their number of
mutual acquaintances in the network, their number of previous
collaborations, and their number of previous collaborators.  We find that
the probability of collaboration is strongly positively correlated with
each of these, and for the latter two that the relationship is close to
linear over a large part of its range.  These results lend strong support
to previously conjectured theories about the way in which networks grow.

\section*{Acknowledgements}
The author thanks L\'aszl\'o Barab\'asi and Duncan Watts for helpful
conversations, and Erzs\'ebet Ravasz for providing an early preprint of
Ref.~\onlinecite{BJNRSV01}.  This work was funded in part by the National
Science Foundation and by Intel Corporation.


\begin{references}
\bibitem{note1}
  This assumption is debatable.  Certainly scientific collaboration
  patterns have changed over the last century, so that $R_m$ probably does
  vary with $t$ on long time-scales.  Whether this variation is visible
  over the course of a single year is unclear.
\end{references}
\end{document}